\documentstyle[epsf,fleqn,aps]{revtex}

\def\Section#1{}

\def\Sv{\vec{S}}

\def\beq{\begin{equation}}
\def\eeq{\end{equation}}
\def\bea{\begin{eqnarray}}
\def\eea{\end{eqnarray}}

\begin{document}
\twocolumn[\hsize\textwidth\columnwidth\hsize\csname
@twocolumnfalse\endcsname
\title{Magnetization plateaus in dimerized spin ladder arrays}

\author{D.C.\ Cabra$^{1,2}$ and M.D.\ Grynberg$^1$}

\address{$^1$Departamento de F\'{\i}sica, Universidad Nacional de La Plata,
C.C.   67, (1900) La Plata, Argentina.\\
$^2$Facultad de Ingenier\'{\i}a, Universidad Nacional de Lomas de Zamora,\\
Cno. de Cintura y Juan XXIII, (1832), Lomas de Zamora, Argentina.}
\maketitle
\date{\today}
\maketitle

\begin{abstract}
We investigate the ground state magnetization plateaus appearing in
spin-$\frac{1}{2}$ two-leg ladders built up from dimerized antiferromagnetic
Heisenberg chains and dimerized zig-zag interchain couplings.
Using both Abelian bosonization and Lanczos methods
we find that the system yields rather unusual plateaus and
exhibits massive and massless phases for specific choices or ``tuning''
of exchange interactions. The relevance of this behavior
in the study of NH$_4$CuCl$_3$ is discussed.

\vspace{10 pt}

Published in Phys. Rev. {\bf B} 62, 337 (2000).
\vspace{-12 pt}
\end{abstract}

\vskip 0.2cm
\vskip2pc]

%_____________________________________________________
\section{Introduction}
%_____________________________________________________

The study of low dimensional antiferromagnets
is playing a major role and bringing new insights in
our current understanding of collective spin fluctuations
\cite{Elbio}.
Part of the fascination of this field is
that in the extreme quantum limit of $S=1/2\,$ there is
a rather complex cross-over between 1-$d$ Bethe ground states
and 2-$d$ long range order. By now it is well established
that, at least within the strong coupling regime, ladders
with an even number of chains exhibit massive
phases with purely short-range spin correlations. In contrast,
odd-chain ladders display in general massless spin excitations
resembling typical 1-$d$ ground states with power-law decaying
correlations. A wealth of experimental investigations have confirmed
these expectations for an increasing number of materials, such
as families of low dimensional cuprates like Sr-Cu-O
and La-Cu-O compounds \cite{Elbio,Hiroi} which can be well described
in terms of spin-1/2
Heisenberg antiferromagnets confined to ladder geometries.

Surprisingly, recent measurements of  magnetization curves in
NH$_4$CuCl$_3$ crystals at high magnetic fields \cite{Tanaka}
revealed rather unexpected features which contrast with the
general behavior of spin excitations expected for these systems
\cite{OYA}-\cite{CHP2}. Specifically, this two-leg ladder $S=1/2$
dimerized compound displays two magnetization plateaus at
one-quarter and three-quarter of the saturation magnetization,
irrespective of the external field direction. These results
confront one of the central issues regarding the condition of
fractional quantization for the appearance of massive spin
excitations or magnetization plateaus under external varying
fields. On general grounds this condition can be shown to be
\cite{OYA}-\cite{CHP2} \beq \label{fractions} p \, N \,S\,( 1 -
\langle M \rangle \,) \in {\cal Z}\,, \eeq where $p,\,$ $N\,$ and
$S\,$ stand respectively for the periodicity of the ground state,
the number of coupled chains and the total spin per site, whereas
$\langle M \rangle$ denotes the system magnetization normalized to
its saturation value. Thus, at sufficiently low temperatures two
coupled spin-1/2 dimerized chains ($p=N=2$) should exhibit, if
any, two plateaus at $\langle M \rangle =\,$ 0 and 1/2. However,
observations  at temperatures down to 0.5$^o$K and magnetic fields
up to 35\,T carried out in \cite{Tanaka} showed no evidence of
these latter plateaus. The situation is rather intriguing since
apart from 3$d$-effects in NH$_4$CuCl$_3$ as well as
low-temperature structural transitions which can not be ruled out,
the magnetization of a number of low dimensional halides and
pnictides, e.g. KCuCl$_3$\,\cite{Tanaka,Tanaka2,KCuCl3},
Cs$_2$CuCl$_4$\,\cite{Coldea}\,, seems to be in fair agreement
with  Eq.\,({\ref{fractions}).

Motivated by these conflicting observations, recent theoretical
studies \cite{Sierra1,CG2} pointed out that two-leg ladder
antiferromagnetic systems can exhibit vanishing spin gaps
depending on the {\it manner} in which the array of coupling
exchanges is realized. The key issue is that the interplay between
massive generating mechanisms such as dimerization
and interchain couplings eventually yield
no cost in energy to create spin excitations under magnetic
fields. Therefore, the system switches from one ground state to
another (short or long ranged), depending in a highly non-trivial
way on the particular choice or ``tuning'' of coupling exchanges.
This is an ubiquitous sign of the importance of quantum
fluctuations of individual spins on the ground state. Depending on
the exchange interactions, fluctuations can manifest themselves
collectively into many possible ground states, particularly in
lower dimensions where their effects are most pronounced.

In this work we further analyze these findings by means of
two independent and complementary techniques such as Abelian bosonization
\cite{OYA}-\cite{CHP2}
and Lanczos diagonalization \cite{Lanczos} of finite systems.
We shall consider  ladders of richer topologies including both
frustrated interactions and dimerization along
the interchain and intrachain couplings. Specifically,
we focus attention on  two dimerized spin-1/2 chains interacting
through an isotropic Hamiltonian of the form
\bea
\nonumber
H = \!\! \sum_{a=1,2,\;n=1}^L \!\! J_n^{(a)} \; \Sv_n^{(a)}\!
\cdot\! \Sv_{n+1}^{(a)}\\
+  J_2 \sum_{n=1}^L \left[ (1-\delta_2) \Sv_{n}^{(1)} \Sv_{n}^{(2)}
\label{H12} + (1+\delta_2) \Sv_{n}^{(1)} \Sv_{n+1}^{(2)} \right]\,
\eea
where the $\Sv_n$ denote spin-$1/2$ operators associated to site $n$.
The array of coupling exchanges
are set as $J_n^{(2)} \equiv J_{n+1}^{(1)}$, and parameterized by
$J_n^{(1)}=J_1\,\left[\,1\,+ (-1)^n \,\delta_1 \,\right]$, say for chain (1).
To maintain purely antiferromagnetic interactions throughout the $2 L$
spins of the ladder with periodic boundary conditions,
the dimerization parameters are kept bounded by
$\vert\, \delta_i \,\vert \le 1\,, i=1,2\,$.  The corresponding zig-zag
structure is schematized in Fig.\, 1.

Besides its theoretical interest \cite{CHP2}, at room temperature
this triangular topology is actually realized both in
NH$_4$CuCl$_3$ \cite{Tanaka} and KCuCl$_3$ \cite{Tanaka,Tanaka2} crystals.
As it was referred to above, their
magnetization curves exhibit quite different features thus, we are especially
interested to include an homogeneous field term of the type
\beq
\label{Hh}
H_h= - \,\frac{h}{2} \,\sum_n \, \left[ S_n^{z (1)} + S_n^{z (2)} \right]\,,
\eeq
so as to elucidate the combined effect of the above kinds of coupling
arrays and applied magnetic fields $h$, say along the $z$ direction.
Also, it is worth pointing out that anisotropic exchanges can be
included straightforwardly in our bosonization procedure
as well as in the numerical analysis.

In studying the ground state regimes of different
exchange parameter sets, two subcases of particular
interest arise immediately \cite{Sierra1,CG2}.
Clearly, by setting $\delta_2 = -1$ (1)
we obtain a non frustrated dimerized system with staggered (plain)
bond alternation as indicated in Fig.\,2.
The analyses given in Ref. \cite{Sierra1,CG2} have shown
that the magnetization behavior of the former
case resembles that of a single dimerized chain, as opposed
to the non-staggered or plain situation ($\delta_2 = 1$\,),
in which a net magnetization plateau shows up at
$\langle M \rangle = 1/2\,$.
It will turn out that our general zig-zag
ladder interpolates continuously between the
above scenarios and yields rather robust plateaus at
$\langle M \rangle = 0\,$ and 1/2.
However, the fine tuning of
exchanges can suppress these massive regimes and furthermore,
for particular subsets of the parameter space, in addition two
magnetization plateaus at $\langle M \rangle = 1/4\,$ and 3/4
emerge simultaneously. This latter issue actually could shed
light on the measurements reported in Ref. \cite{Tanaka}.

The layout of the paper is organized as follows. In Section 2, we
recast the low-energy excitations involved in each chain of
Eq.\,(\ref{H12}) in terms of a conformal field theory of a free
bosonic field compactified at a magnetization dependent radius.
The dimerization along chains ($\delta_1$) and zig-zag
interactions ($\delta_2$) render the bosonization approach
particularly suitable to examine respectively weak ($J_2/J_1,\,
\vert \delta_1 \vert \ll 1\,$) and strong ($J_1/J_2,\, \vert
\delta_2 \vert \ll 1\,$) coupling regimes. They are treated in
turn in Section 2 A and 2 B.\, Section 3 complements the
magnetization behavior conjectured by the analytic approach in a
variety of non-perturbative scenarios. An exact numerical
treatment of magnetization contours for finite systems up to $2L =
24$ spins is given using a recursion-type Lanczos algorithm
\cite{Lanczos} applied on each magnetization subspace.  Standard
extrapolation procedures\cite{Gutt} to the thermodynamic limit
then enable an independent test of the results obtained via
bosonization techniques. We end the paper with Section 4 which
contains our conclusions, along with some remarks on pros
and limitations of the present work.

%________________________________________________________________
\section{Abelian bosonization}
%________________________________________________________________

Following a recent analysis discussed as in Refs.\
\cite{OYA}-\cite{CHP2}, we will apply the by now standard
method of Abelian bosonization at Hamiltonian (\ref{H12}). In this
formalism an antiferromagnetic homogeneous chain is described by a
compactified free bosonic field $\phi^{(a)}$ whose dynamics is
governed by \beq H^{(a)} = {1 \over 2} \int dx \left( v K
(\partial_x \tilde{\phi}^{(a)})^2 + {v \over K} (\partial_x
\phi^{(a)})^2 \right)
\label{Hambos}
\eeq
The dual field $\tilde{\phi}^{(a)}$ is defined as usual
$\Pi = \partial_x \tilde{\phi}^{(a)}$. $v$ is the Fermi velocity and
the Luttinger
constant $K$, which is a function of the magnetization $\langle
M^{(a)} \rangle$ and an eventual $XXZ$ anisotropy $\Delta$, governs the
conformal dimensions of the bosonic vertex operators and can be
obtained exactly from the Bethe ansatz solution of the $XXZ$ chain
(see e.g.\ \cite{CHP} for a detailed summary). It is related to
the compactification radius $R$ of \cite{CHP} by $K^{-1} = 2 \pi R^2$.

In terms of these fields, the spin operators read
\bea
S_{x}^{z,(a)} \sim {1 \over \sqrt{2\pi}} \partial_x \phi^{(a)} +\
\ \ \ \ \ \ \ \ \ \ \ \ \ \ \ \ \ \ \nonumber \\ \ \ \ \ \ \ \ \ a
: \cos(2 k_F^i x + \sqrt{2 \pi} \phi^{(a)}): + \frac{\langle
M^{(a)}\rangle}{2} \, , \label{sz} \\ S_{x}^{\pm,(a)} \sim (-1)^x
:{\rm e}^{\pm i\sqrt{2\pi} \tilde{\phi}^{(a)}} \ \ \ \ \ \ \ \ \ \
\ \ \ \nonumber \\ \left( b \cos(2 k_F^{(a)} x + \sqrt{2 \pi}
\phi^{(a)}) + c \right) : \, ,
\label{s+}
\eea
where the colons denote normal ordering with respect to the ground
state with magnetization $\langle M^{(a)}\rangle$. The Fermi momentum
$k_F^{(a)}$ is related to the magnetization of the $a^{\rm th}$ chain as
$k_F^{(a)} = (1-\langle M^{(a)} \rangle )\pi/2$. The effect of an
$XXZ$ anisotropy $\Delta$ and/or the external magnetic field is
then to modify the scaling dimensions of the physical fields
through $K = K(\langle M^{(a)}\rangle, \Delta)$. The magnetization
also modifies the
commensurability properties of the spin operators
through $k_F$, as can be seen
from  (\ref{sz}), (\ref{s+}).
The non-universal constants $a$, $b$
and $c$ can be in general computed numerically (see e.g.\
\cite{HF}, for the case of zero magnetic field) and in particular
the constant $c$ has been obtained exactly in \cite{LZ}.

%________________________________________________________________________
Notice that the inclusion of an $XXZ$ anisotropy in our study is
not only motivated to generalize the analysis but primarily  by the fact
that, for non-zero magnetization, the $SU(2)$ symmetry is broken from the
beginning. As we shall see in  Section 2 A,
the particular $SU(2)$ symmetric case, ($\Delta = 1$,
$\langle M \rangle = 0$), has to be analyzed differently since our analysis
breaks explicitly this symmetry. We address the reader to Ref.\cite{NGE}
where the full symmetric was analyzed using a formulation in terms
of Majorana fermions.
%________________________________________________________________________

%____________________________________________
\subsection{Weak interchain coupling regime}
%____________________________________________

Here we take $\alpha \equiv J_2/J_1 \ll 1$ and $\delta_1 \ll 1$.
In this regime $\vert \delta_2 \vert \ll 1$ corresponds to a weakly coupled
two-leg zig-zag ladder made up of dimerized chains, whereas for
$\vert \delta_2  \vert \to 1\,$ the system approaches
a staggered ($\delta_2 = -1$) or plain ($\delta_2 = 1$)
ladder dimerized as in Fig.\ 2(a)
and Fig.\ 2 (b) respectively.

Using Eqs. (\ref{sz}, \ref{s+}), the low energy Hamiltonian can be written
in this regime as
\bea
&\phantom{+}&
H_{int}^{(\alpha \ll 1)} \approx  \lambda_1 \sum_x
\partial_x\phi^{(1)}\partial_x\phi^{(2)} \nonumber\\
&+& \lambda_2
\sum_x \cos\left[4k_F x+\sqrt{2 \pi}(\phi^{(1)}+\phi^{(2)})\right]
 \nonumber\\
&+& \lambda_3 \sum_x \cos\left[\sqrt{2 \pi}(\phi^{(1)}-\phi^{(2)})\right]
\nonumber\\
&+&\lambda_4\sum_x \cos\left[\sqrt{2\pi}
(\tilde\phi^{(1)}-\tilde\phi^{(2)})\right]
\nonumber\\
&+& \lambda_5 \sum_{x,(a)} (-1)^{x+a} \cos\left[2k_F\, (x+1/2)
\phantom{\sqrt{2 \pi}}   \right.\nonumber\\
&+&\left.
\sqrt{2 \pi}\,\phi^{(a)}(x)\right]\nonumber\\
\label{Hint}
&+& \lambda_6
\sum_{(a)=1}^2 \sum_x  \cos\left[2k_F x + \sqrt{2 \pi}\,\phi^{(a)}(x)\right]
\eea
where
\beq
\hspace{-0.8cm}
 \lambda_j / \alpha\, \propto \cases{\Delta_2\,,\:\: j = 1\,,\cr
\Delta_2 \left[(1-\delta_2)\! +\! (1 + \delta_2)
\cos (2k_F)\right]\,, j=2,3\,,\cr
 \delta_2\,,\:\:\:\: j = 4\,,\cr
 \delta_1\,,\:\:\:\: j = 5\,,\cr
 \langle M \rangle\,, j = 6\,.\cr}
\eeq
In the last two expressions the proportionality factors
are non-universal functions of the $XXZ$ anisotropy $\Delta_2$,
the magnetization $\langle M \rangle$ and $\delta_2$.

%------------------------------------------------------------------
In the above equation we have suppressed marginal terms, due to
the presence of more relevant interactions. Also in the case of
non-dimerized zig-zag interactions, there are parity breaking
terms, discovered in \cite{NGE}, which should be analyzed
differently since they have non-zero conformal spin. However, by
explicitly including these latter terms in the RG computations, it
can be shown  that they do not change the conclusions presented
here. The $SU(2)$ symmetric case ($\Delta = 1$, $\langle M \rangle
= 0$) has been studied in \cite{NGE} for the case of zero zig-zag
dimerization.

%------------------------------------------------------------------
Let us analyze the structure of the magnetization curves predicted
by this effective Hamiltonian. The relative field $\phi_- = \phi_1
- \phi_2$ is always massive due to the fact that the relevant
perturbation terms $\lambda_{3,4}$ are always commensurate. The
analysis for the diagonal field $\phi_+ = \phi_1 + \phi_2$ is more
subtle so we will consider separately different values for the
magnetization. The possible values of the magnetization at which
plateaus can appear are given by the general
expression (\ref{fractions}) \cite{OYA,CHP}.
However, due to the presence of frustration in certain
region of the parameter space, possibly  Eq. (\ref{fractions})
should include an extra factor 2 to account for
a possible enhancement in the periodicity of the ground state.
This restricts the set of possible magnetization plateaus to the
values $\langle M\rangle =0,1/4,1/2,3/4$ (apart from saturation).

For $\langle M\rangle =0$, ({\it i.e.} $k_F=\pi/2$), the
perturbation $\lambda_2$ becomes commensurate and hence opens a
gap for the diagonal field $\phi_+$. One could however argue that
taking into account other (radiative) corrections (coming mainly
from $\lambda_5$ and $\lambda_6$) one could close the $\phi_+$
gap, simply by making the amplitude of this perturbation term to
vanish. This naive analysis has been confirmed by numerical
computations\cite{CG2}, showing that there exists a whole curve in
the parameter space where this indeed happens.
In particular, for $\delta_2 = -1$ the critical line turns out to be
\beq
\label{critical}
\frac{J_2}{J_1} \propto \delta_1^2\,.
\eeq
This phenomenon was originally suggested in \cite{Sierra1} using
non-linear sigma model techniques and has been the also studied
numerically in \cite{Sierra2,Austr}.

For $\langle M\rangle =1/2$, ({\it i.e.} $k_F=\pi/4$), the plateau
can open due to the radiatively generated (relevant) terms,
(coming from $\lambda_5$ and $\lambda_6$), which are of the form
\beq
\hspace{-0.5cm}
\alpha^2 \delta_1\langle M\rangle
f(\delta_1,\delta_2,\Delta_2) (-1)^x
\cos(4k_F x + \sqrt{2\pi}\phi_+),
\label{1/2}
\eeq
where $f(\delta_1,\delta_2,\Delta_2)$ vanishes for the case of
$\delta_2= -1\,$, as already pointed out in \cite{CG2}.
As it will be shown in Section 3, possibly there is
another point at which the $\langle M\rangle =1/2$ plateau closes.
The existence of such effects can be predicted using the
bosonization formalism, but
the precise location of this point, however, cannot be obtained
due to the presence of non-universal constants in Eqs.\
(\ref{sz}), (\ref{s+}).

The case of $\langle M\rangle =1/4,3/4$, ({\it i.e.}
$k_F=3\pi/8,\pi/8$), is less clear since the commensurate
operators that can be generated to open this plateau are
irrelevant. One possible candidate is the operator $(-1)^x
\cos(8k_F x + 2\sqrt{2\pi}\phi_+)$ which is irrelevant.
One can speculate whether this
operator can become relevant in some region of the parameter space
by taking into account the $\lambda_1$ perturbation which changes
the compactification radius of the $\phi_+$ field, and hence the
dimensions of the perturbation terms. However, in first
approximation the above mentioned operator would become
relevant for values of the couplings far from the region where our
approach can be considered valid.

\vskip 1cm

%______________________________________________
\subsection{Strong interchain coupling regime}
%______________________________________________

Let us consider now the opposite regime, {\it i.e.} $\alpha \gg 1\,$.
In this regime it is convenient to rewrite the two-leg zig-zag
ladder Hamiltonian as that of a single chain with
alternating intrachain
coupling $J_2 (1\pm \delta_2)$ and next-nearest-neighbor (NNN)
interactions $J_1(1\pm \delta_1)$ but alternating every two sites.
Namely, NNN spins at $(4n, 4n+2)$ and $(4n+1, 4n+3)$ pair locations
are coupled by $J_1(1-\delta_1)\,$,
whereas spins at $(4n-2, 4n)$\, and \, $(4n-1, 4n+1)$
interact through $J_1(1+\delta_1)\,$. This is illustrated
in Fig. 3.

Using the same approach as described above, in the present case we
get an effective Hamiltonian for a {\it single} bosonic field
perturbed by the following terms

\bea
&\phantom{+}&
H_{int}^{(\alpha \gg 1)} \approx
 \sum_x \gamma_1(x) \left(\partial_x \phi \right)^2 \nonumber\\
&+&  \sum_x \gamma_2(x) \cos\left[2k_F x +
\sqrt{2 \pi}\,\phi(x)\right] \nonumber\\
&+& \gamma_3 \sum_{x}(-1)^x \cos\left[2k_F (x+1/2) +
\sqrt{2 \pi}\,\phi(x)\right] \nonumber\\
&+&  \sum_x \gamma_4(x) \cos\left[ 4k_F x +
2\sqrt{2 \pi}\,\phi(x)\right]\ .
\label{NNN}
\eea
Here $\gamma_{1,2,4}(x)$ are proportional to $\delta_1$ and
$\gamma_3$ to $\delta_2$. The terms proportional to $\delta_1$
have an extra modulation due to the alternation of the
NNN couplings  referred to above, which hereafter we call
``two-by-two" modulation.

For $\delta_1 =0$ and $\langle M \rangle =0$ only the last term
is commensurate and it has dimension $2$ in the isotropic case ($\Delta =1$).
This term is the one responsible for the opening of the gap
at zero magnetization in the NNN antiferromagnetic Heisenberg
chain, where, as it is well known, the gap opens at a critical value
of the NNN coupling through a Kosterlitz Thouless (KT)
transition. Its effect would have been the same in the
present case, but due to the presence of the dimerization
$\delta_1$ along the chain, a gap will always be present. This will
be corroborated in Section 3.

For $\langle M \rangle =1/2$, the perturbation $\gamma_2$
survives the continuum limit due to the extra ``two-by-two"
alternating factor, and will hence be responsible for the
plateau at this value of the magnetization.

The case of $\langle M\rangle =1/4,3/4$ is again more subtle in
this limit, and it can be seen that the operator that could be
responsible for the appearance of these plateaus is generated from
a combined effect of the chain dimerization ($(-1)^x \delta_2$) and the
{\it two-by-two alternating part} of the NNN exchange ($\pm \delta_1 J_1$).
The operator generated through this mechanism is proportional to
$\cos(2\sqrt{2 \pi}\,\phi(x))$, which is irrelevant, and
could become relevant at certain critical value of the
coupling ($\propto J_1 \delta_1 \delta_2$). Again, this KT point
is not reachable within our perturbative approach.

%_____________________________________________________
\section{Numerical analysis of finite systems}
%_____________________________________________________

To enable an independent check of the magnetization scenario
obtained within the bosonization approach, we now turn to a
numerical finite-size analysis of the original ladder Hamiltonian
(\ref{H12}). A number of numerical studies of triangular
ladders (or equivalently, of Heisenberg chain with NN exchanges),
have been reported already (\cite{CHP2,Ger} and references therein).
However, the effect of
dimerization along both interchain and intrachain couplings, which is
crucial for the appearance of non-trivial magnetization plateaus,
yet requires further numerical efforts.

We focus attention on the ground state energy obtained from
an exact diagonalization of finite systems via a recursion type
Lanczos algorithm \cite{Lanczos} applied on each magnetization subspace
with $S^z = \left\{0,\,1,\,...\,, L\right\}$. Since the magnetic field
considered in Eq.\,(\ref{Hh}) is coupled to the conservation of
$S^z/L\,= \langle M \rangle$, we can readily relate the energy
per spin $e_h$ at finite fields to those at $h=0\,$ just by taking
$e_h \equiv e_0 - h\,\langle M \rangle$. Thus, all results addressed
below were obtained from computations with $h = 0\,$.
Also, to avoid unwanted  effects introduced by both the ladder
topology and periodic boundary conditions, even multiples of ladder
lengths up to $L = 12\,$ spins were taken throughout.

The huge dimensionality of the spaces involved, growing
as $2L \choose L$, constrained us to use in the
heaviest situations, i.e. $S^z = 0,\,1\,$, at most seven Lanczos
vectors per tridiagonalization cycle \cite{comment}. Nevertheless,
the numerical accuracy was kept
bounded by $10^{-7}\, h/J_1$ employing typically up to forty cycles
of recursion. On the other hand, the rather small number
of Lanczos vectors used in the computations
allowed for an efficient
management of a complete, vector by vector reorthogonalization.
As is known \cite{Lanczos}, this latter procedure becomes crucial
to avoid the emergence of spurious eigenvalues caused by machine
rounding errors which tend to build up exponentially with the number
of iterations, no matter what precision is used.
In what follows we limit our analysis to isotropic coupling exchanges,
though preliminary calculations including anisotropy in the field
direction yield qualitatively similar results.

We begin by examining the validity
of Eq.\,(\ref{fractions}) and test the ``fine tuning'' effects
conjectured within the bosonization analysis.
Upon setting $\delta_1 = 0.7$, $\delta_2 = 0.475$
and $J_2/J_1 = 1$, the application of the Lanczos procedure
to the triangular ladder Hamiltonian  (\ref{H12}) yields
a quite unusual behavior. Specifically,
in addition to the typical $\langle M \rangle = 0\,$
and 1/2 plateaus, two massive phases appear simultaneously
at $\langle M \rangle = 1/4\,$ and 3/4 as is shown
in Fig. 4. Rather unexpectedly,
at this point of the parameter space the spin ladder seems to decouple
into a quasi-four level system (see the region near the
vertical lines of Fig. 5).
In principle this renders
size effects less pronounced, possibly due to some enhanced
or hidden symmetry whose origin is yet difficult to elucidate.
Nevertheless,  our results provide a strong numerical evidence
indicating  that  zig-zag structures can yield non trivial
magnetization plateaus such as those observed in NH$_4$CuCl$_3$
crystals~\cite{Tanaka}.

Interesting as it is, we are actually motivated to obtain
the {\it whole} magnetization curve of this latter compound. Therefore,
we turn to the scanning of the exchange parameter space for different
regimes. Figs. 5(a), 5(b) and 5(c) display respectively
typical magnetization contours or ``phase diagrams''
for weak ($J_2/J_1 = 0.5$, $\delta_1 = 0.9\,$),
intermediate ($J_2/J_1 = 1$, $\delta_1 = 0.7\,$),
and strong ($J_2/J_1 = 5 $, $\delta_1 = 0.5\,$),
coupling regimes. This is a compact form of
representing conventional magnetization curves for a wide range of
interchain couplings. Here each line is associated to
successive values of $\langle M \rangle$  which increase monotonically
with the applied field $h$. For example, the magnetization plateaus
of Fig. 4 are contained completely within  the vertical line of
Fig. 5(b). In general, we found that the $\langle M \rangle = 0\,$
and 1/2 plateaus remain robust in a variety of scenarios, though their
widths can be eventually ``fine tuned'' to yield massless gaps.

In studying the mass gap extrapolation towards their
thermodynamic limits (i.e. the energy gap to create
an excitation of total spin $S=1\,$ as $L \to \infty\,$),
we fitted the whole set of finite-size results for
$4 \le L \le 12$ ($L$ even), using a variety of
standard procedures. These range from linear to logarithmic and van den
Broeck-Schwartz type methodologies of convergence \cite{Gutt}, which
basically yield analogous results with at least two significant digits.
We draw the reader's attention to panels 6(a)--(b),\,\, 6(c)--(d),\,\, and
6(e)--(f)\, in which we display respectively
gap extrapolations around $\langle M \rangle = 1/2\,$ and 0
corresponding to the ground state regimes exhibited in Figs. 5(a)--(c).
Similar gap extrapolations for
$\langle M \rangle = 1/4\,$ and 3/4 beyond the ``symmetry''or
finite size collapse region  denoted by the vertical lines
of Figs. 5(a) and 5(b),  would be unreliable given
the scarcity of available data. In our case, this is translated in
the availability of matching sizes, namely $L = 4,\,8,\,12$, already
constrained by the studied values of $\langle M \rangle\,$.
Thus, it remains unclear whether or not empty wide ``bands''
or plateau regions for $\langle M \rangle = 1/4\,$ and 3/4
could be actually present in Figs. 5(a)--(c).

To complement the analysis of finite (vanishing)
$\langle M \rangle = 1/2\,$ gaps
for $\delta_2 = 1$ (-1) given in Section 2 A,  we see that
their widths remain stable upon setting $\delta_2  < 1\,$,
($\delta_2  > -1\,$). Moreover, the wide minima
of Figs. 6 (a), (c), and (e) suggest an infinitely continuous
ground state transition at $\delta_2 = -1$, which  is in line with
the KT singularity conjectured in Section 2 B. In fact,
our data  strongly support this picture for a variety of coupling
regimes, at least within
the region $\delta_2 \in (-0.9\,,\,-0.25 )$ where
finite size effects are less pronounced. This can be observed
from the semi-log representation of the data displayed in Fig. 7.
Thus finally, the triangular ladder turns out to
interpolate smoothly between the staggered ($\delta_2 = -1$\,)
and plain ($\delta_2 = 1$\,) dimerization arrays of
the square ladders studied in \cite{CG2}.

%________________________________________________________
\section{Conclusions}
%________________________________________________________

In this work we have studied the magnetization
phase diagram of a two leg zig-zag ladder with dimerization both
along the legs and the zig-zag coupling, by means of
Abelian bosonization methods complemented by Lanczos diagonalization
of finite clusters up to 24 spins.
From the bosonization analysis we conclude that
the $\langle M \rangle = 0$ plateau is robust and it is present
in the full parameter space, except on a certain zero-measure
set. We have confirmed numerically these expectations
and obtained the above massless excitations as displayed
in the lowermost panels 6(d) and (f).

Spin excitations around $\langle M \rangle =1/2$,
also turn out to be massive and robust as they
show up in the whole parameter space,
except in the limit $\delta_2 \rightarrow -1$ and for a  certain curve
$J_2/J_1 = f(\delta_1,\delta_2)$. This latter feature
is observed in Figs.\ 6(a),(c) for weak and intermediate coupling
regimes near $\delta_2 =0$ and around
$\delta_2 \sim 0.5 $ for strong coupling regions
as shown in Fig.\ 6(e).

Regarding the issue of $\langle M \rangle =1/4, 3/4$
plateaus, they are observed only within
a fine-tuned region, possibly bearing an enhanced symmetry
of the Hamiltonian (see e.g. Figs. 4 and 5). Their
appearance is hard to predict using Abelian bosonization
techniques.

Finally, though there are intermediate values of $\delta_2$ capable of
closing gaps around both  $\langle M \rangle = 0\,$ and 1/2,
after scanning a
representative set of the parameter space, we could not find
any evidence of {\it common} closing points so as to explain
the suppression of these plateaus at a time in NH$_4$CuCl$_3$ crystals.
In principle, this would rule out the ladder Hamiltonian (\ref{H12})
as a suitable model to account completely for the experiments
reported in \cite{Tanaka}. Nonetheless, we trust that its success to
describe the {\it simultaneous} emergence of rather unusual
plateaus at $\langle M \rangle = 1/4\,$ and 3/4 (at least near
the data collapse region of Fig. 5), will make our system
worth to consider as an antecedent for future studies
in that direction.

%________________________________________________________
\section*{Acknowledgments}
%________________________________________________________

We appreciate fruitful discussions with  A. Honecker and
P.\ Pujol. The authors acknowledge financial support
of CONICET and Fundaci\'on Antorchas. D.C.C.\ acknowledges financial
support from ANPCyT (under grant
No.\ 03-00000-02249)

%____________ Postscript Files_____________________
\newpage

%_________________ Figure 1 _____________________

\begin{figure}
\hbox{%
\epsfxsize=3.1in
\hspace{1cm}
\epsffile{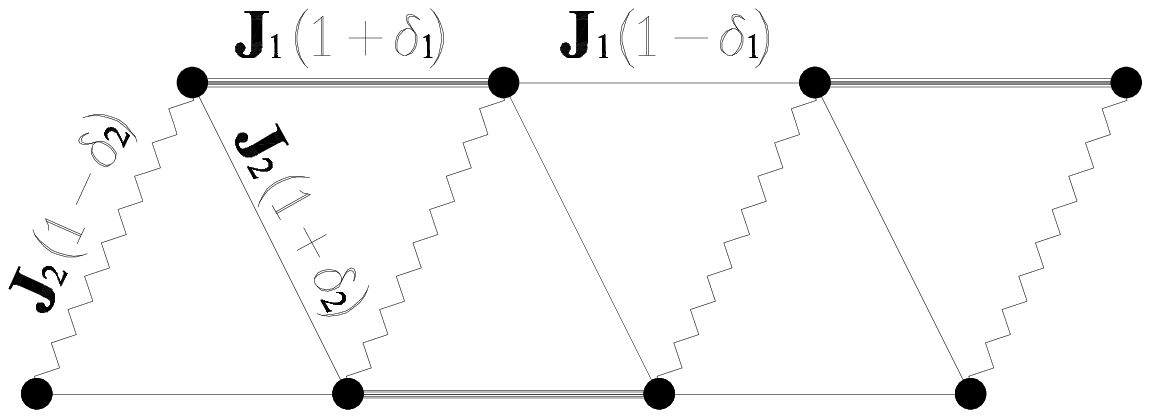}}
\vspace{0.2cm}
\caption{Schematic view of doubly alternating zig-zag ladders
showing both interchain ($J_1$), and intrachain ($J_2$) exchange couplings
along with their respective dimerization parameters $\delta_1$ and
$\delta_2$\,.}
\end{figure}

\vskip 1cm
%_________________ Figure 2 _____________________

\begin{figure}
\hbox{%
\epsfxsize=4.3in
\hspace{-0.8cm}
\epsffile{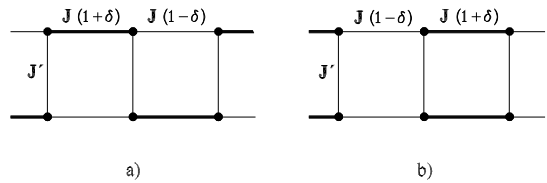}}
\vspace{-1.5cm}
\caption{Representation of non-frustrated ladders
with dimerized chain bonds $J_1 (1\pm \delta_1) \equiv J (1\pm \delta)$
and interchain coupling $J'= 2\,J_2$,  obtained by setting
(a) $\delta_2 = -1\,$ (staggered dimerization)  and, (b) $\delta_2 = 1\,$
(plain dimerization).}
\end{figure}
\vskip 0.7cm

%_________________ Figure 3 _____________________

\begin{figure}
\hbox{%
\epsfxsize=3.9in
\hspace{-1.2cm}
\epsffile{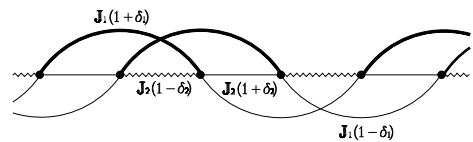}}
\vspace{-1cm}
\caption{Chain representation of dimerized zig-zag ladder}
\end{figure}

\newpage

%_________________ Figure 4 _____________________

\begin{figure}
\hbox{%
\epsfxsize=2.9in
\vspace{2cm}
\hspace{-0.5cm}
\epsffile{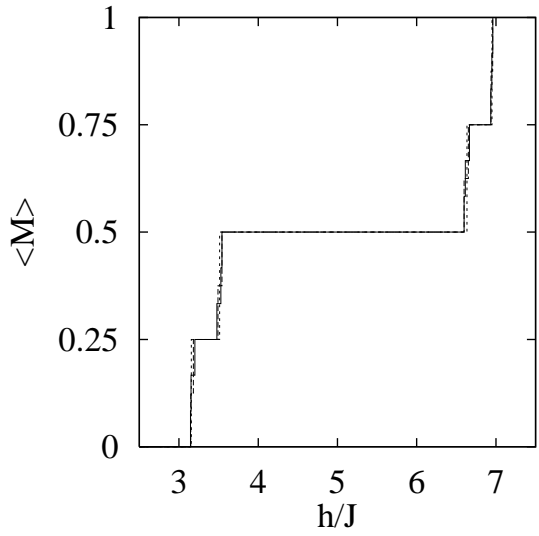}}
\vspace{-1.5cm}
\caption{Magnetization curves of dimerized
zig-zag ladders for $\delta_1 = 0.7$, $\delta_2 = 0.475$
and $J = J_1 =J_2$. Solid, dashed and short dashed lines denote
respectively results for $L=$ 12,8 and 4.}
\end{figure}

\vskip -3.1cm

%_________________ Fig. 5 (a) _____________________
\begin{figure}
\hbox{%
\epsfxsize=4in
\hspace{-1.cm}
\epsffile{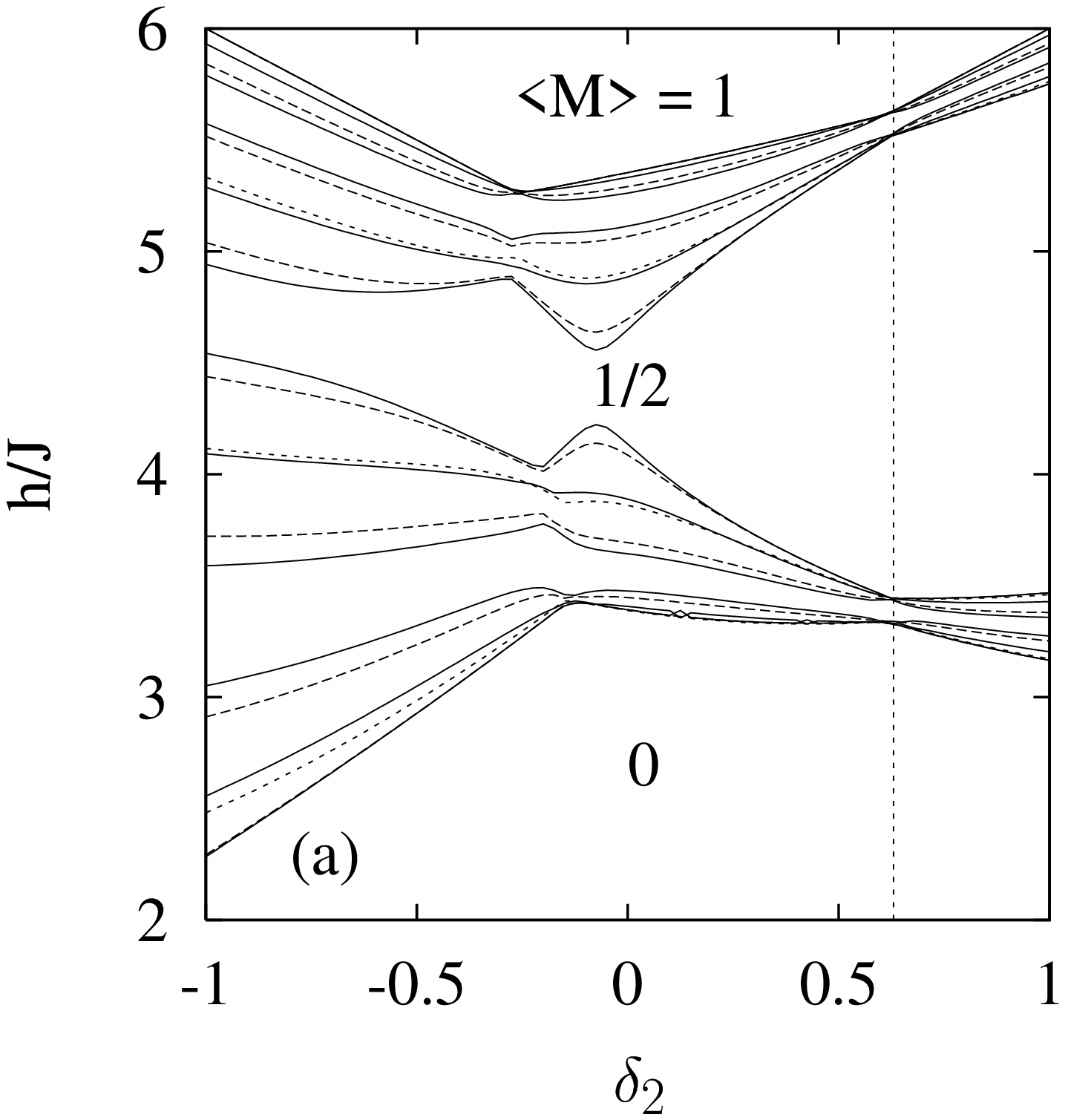}}
\end{figure}

\newpage
%_________________ Fig. 5 (b) _____________________
\hbox{%
\vspace{-3.8cm}}

\begin{figure}
\hbox{%
\epsfxsize=4in
\hspace{-1.cm}
\epsffile{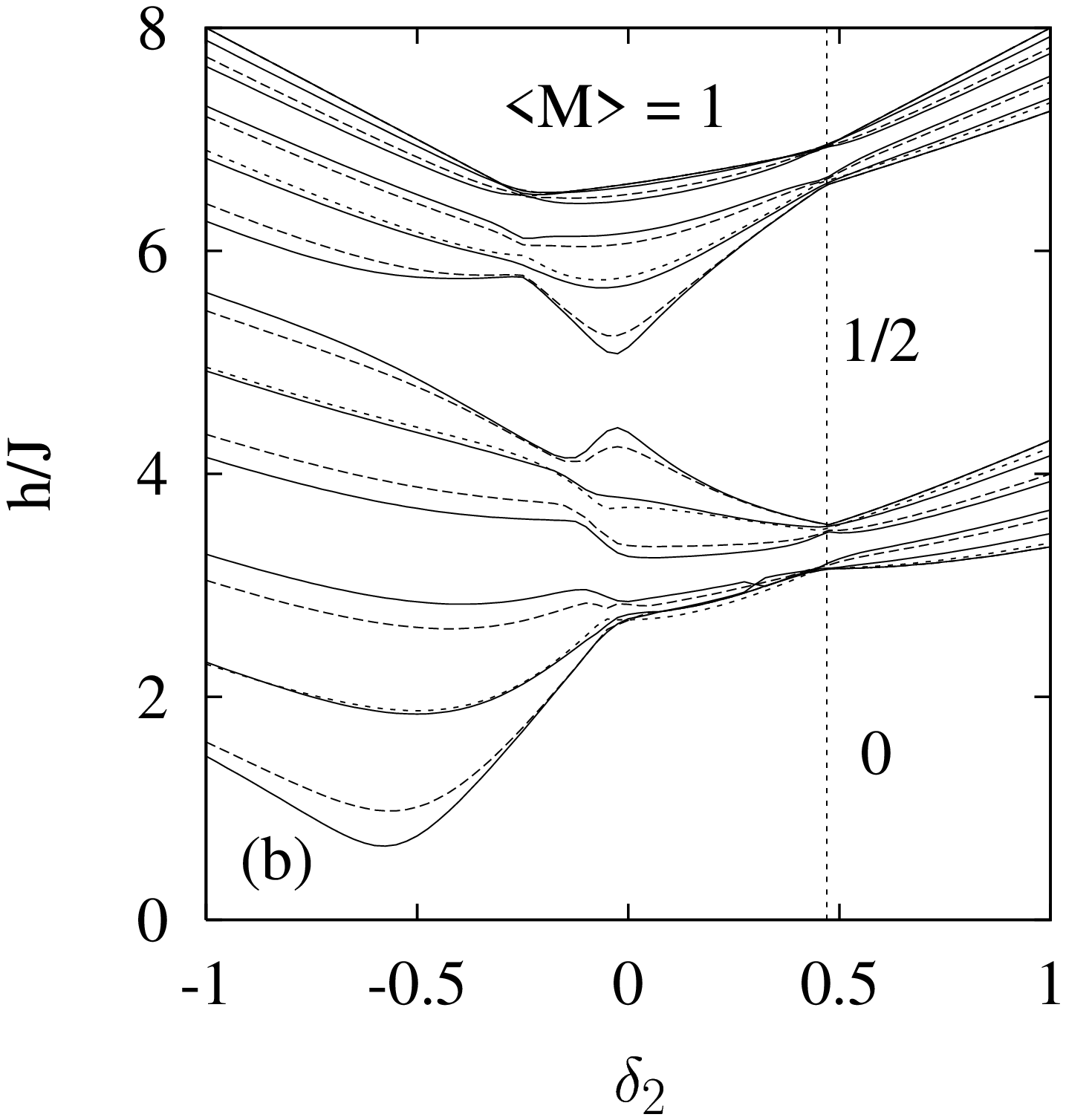}}
\vspace{-3.2cm}
\end{figure}

%_________________ Fig. 5 (c) _____________________
\hbox{%
\vspace{-2.1cm}}

\begin{figure}
\hbox{%
\epsfxsize=4in
\hspace{-1cm}
\epsffile{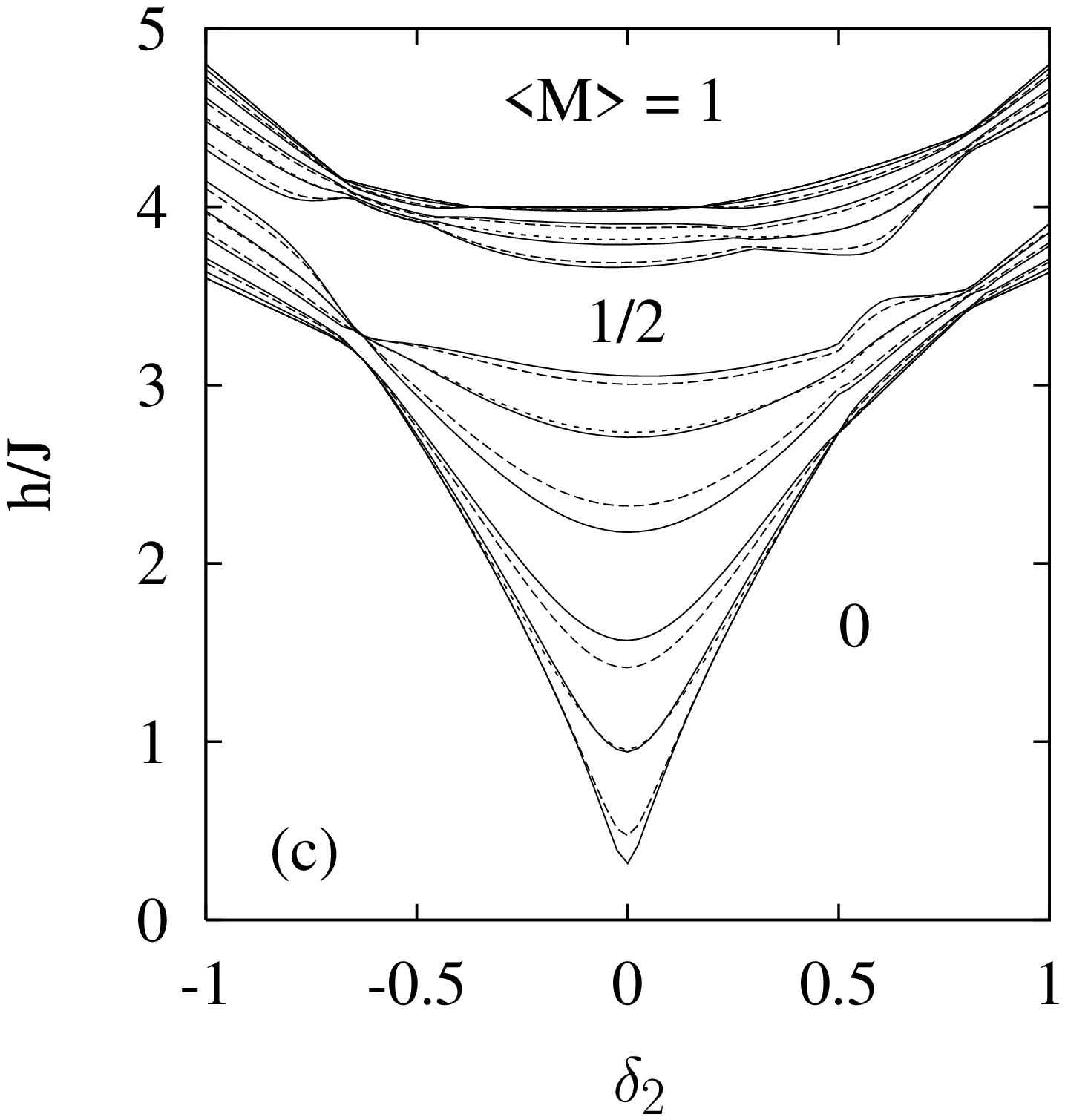}}
\vspace{-3.5cm}
\caption{Critical fields of dimerized zig-zag ladders
for (a) $J_2/J_1 = 0.5$ with $\delta_1 = 0.9\,$, (b)
$J_2/J_1 = 1$ with $\delta_1 = 0.7\,$, and, (c)
$J_2/J_1 = 5 $ with $\delta_1 = 0.5\,$.
Solid, dashed and short dashed lines denote respectively
the results for $L=$ 12,8 and 4, whereas $J = {\rm max}\:
\{J_1,\,J_2\}\,$. Vertical lines in (a) and (b)
indicate regions where 1/4 and 3/4 plateaus emerge at a time.}
\end{figure}

\newpage

%_________________ Fig. 6 a , b _____________________
\hbox{%
\vspace{-3cm}}
\begin{figure}
\hbox{%
\epsfxsize=4.1in
\hspace{-1.2cm}
\epsffile{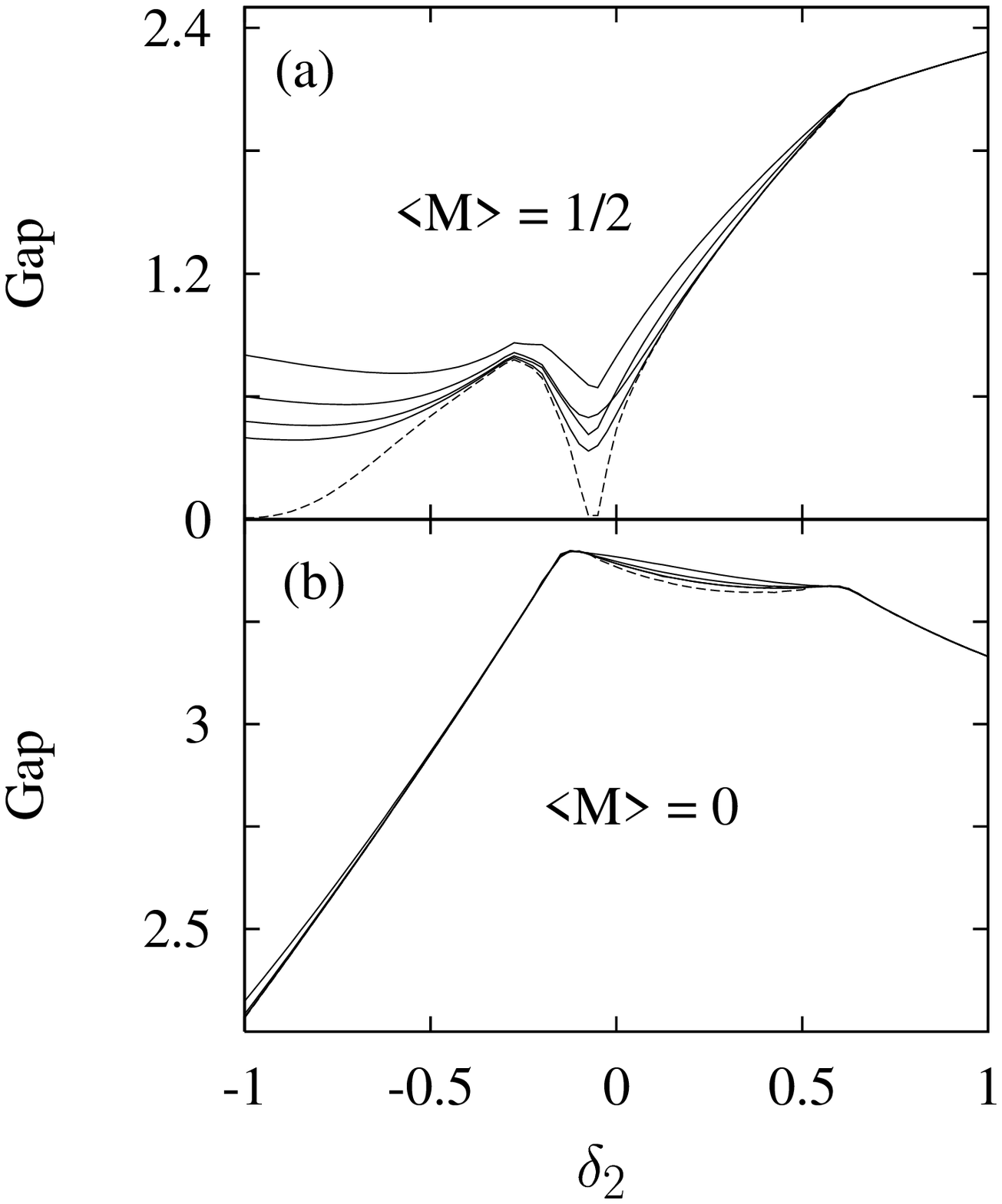}}
\end{figure}

%_________________ Fig. 6 c , d _____________________

\hbox{%
\vspace{-5.5cm}}

\begin{figure}
\hbox{%
\epsfxsize=4.1in
\hspace{-1.2cm}
\epsffile{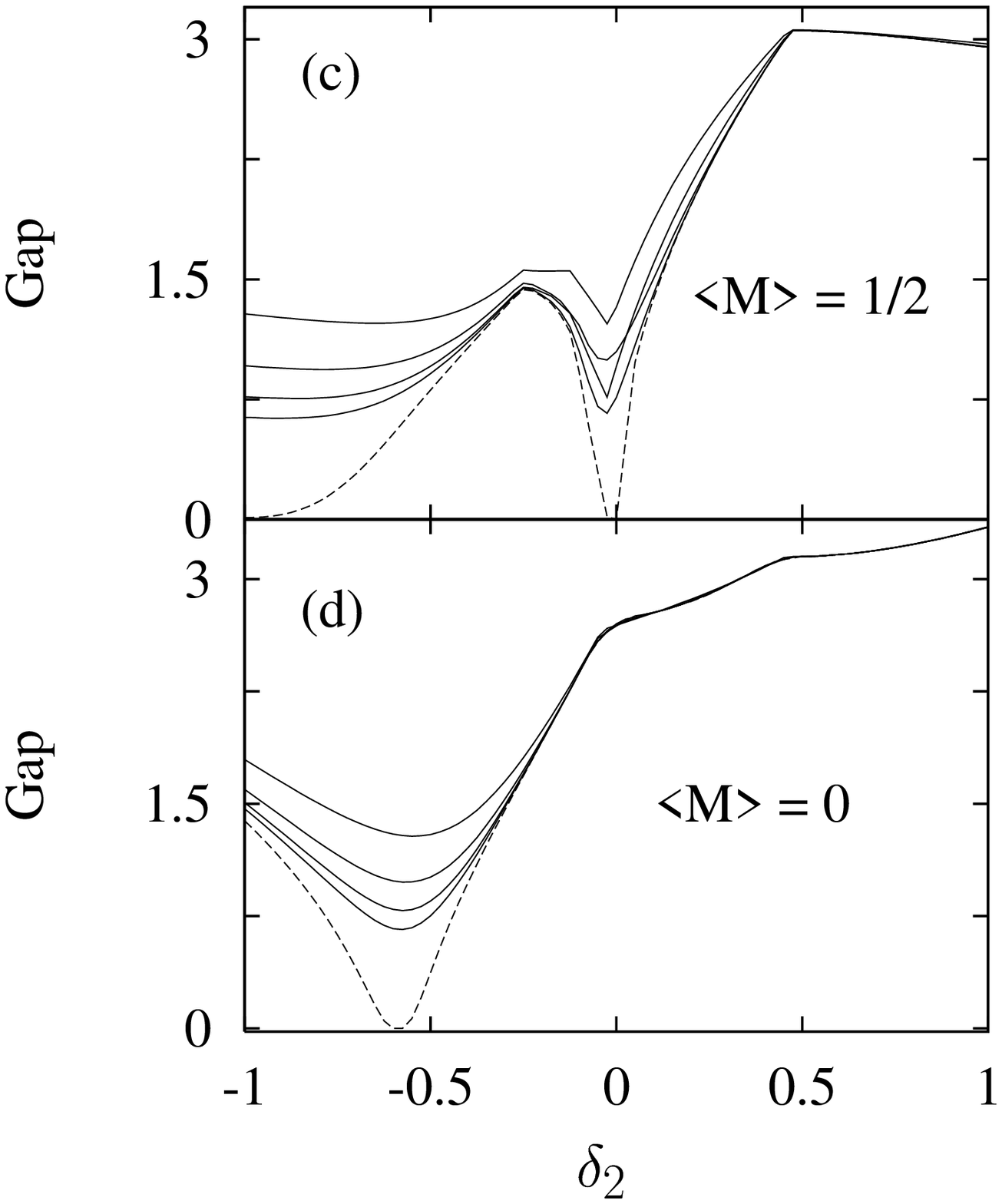}}
\end{figure}

\newpage

%_________________ Fig. 6 e , f  _____________________
\hbox{%
\vspace{-3cm}}

\begin{figure}
\hbox{%
\epsfxsize=4.1in
\hspace{-1.2cm}
\epsffile{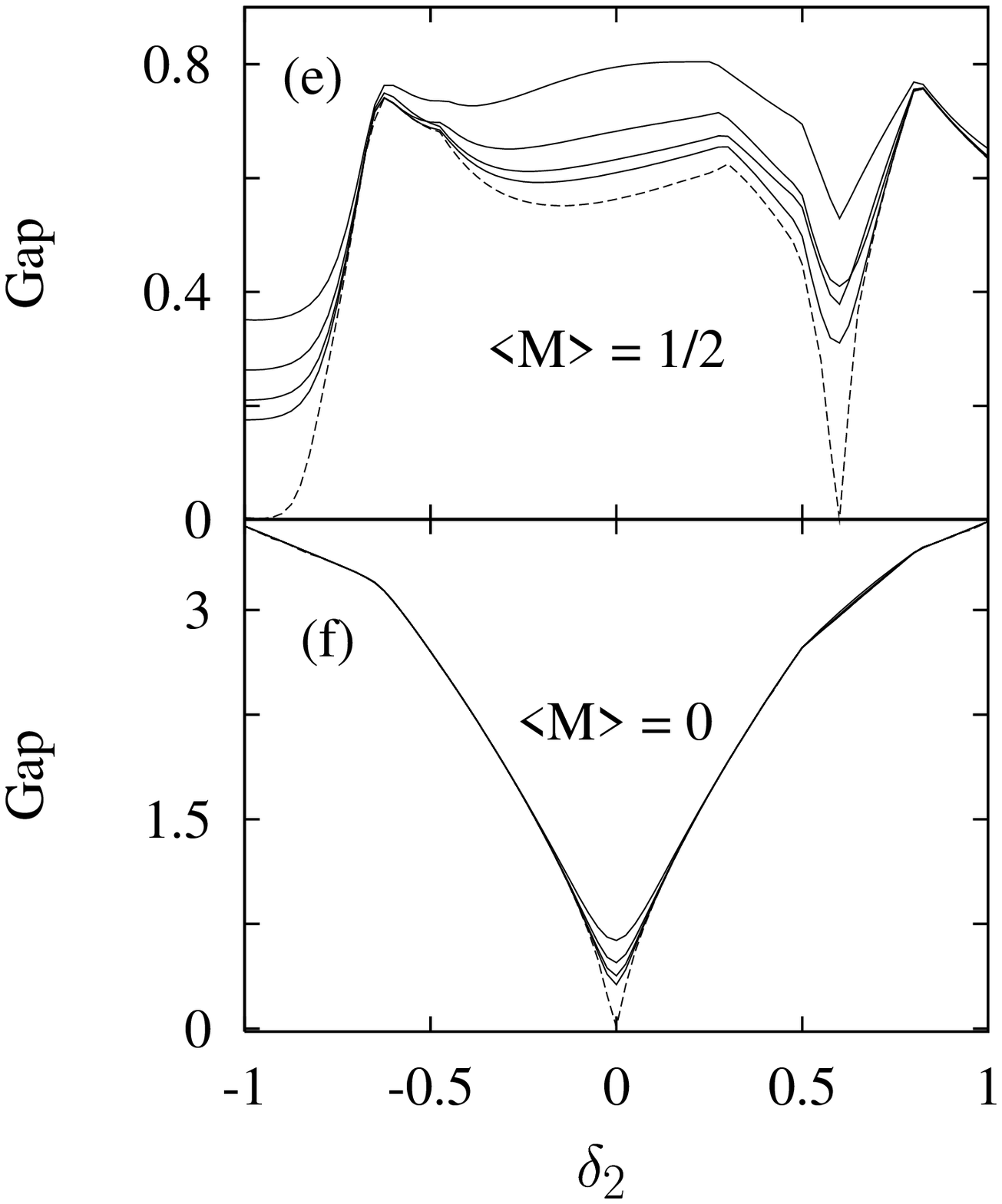}}
\vspace{-2.7cm}
\caption{Width of magnetization plateaus for $\langle M \rangle = 1/2$
in (a), (c), (e) and,  $\langle M \rangle = 0$ in (b), (d), (f).
Data of first, second and third panel refer respectively to
$J_2/J_1 = 0.5$ with $\delta_1 = 0.9\,$,
$J_2/J_1 = 1$ with  $\delta_1 = 0.7\,$ and,
$J_2/J_1 = 5 $ with  $\delta_1 = 0.5\,$.
Solid lines in descending order indicate  respectively
results for $L = 6, 8, 10\,$ and 12, whereas lowermost dashed curves
[slightly visible in (b) and (f)\,]
denote gap extrapolation to the thermodynamic limit.}
\end{figure}

%_________________ Figure 7 _____________________

\begin{figure}
\hbox{%
\epsfxsize=4.5in
\vspace{2cm}
\hspace{-1.7cm}
\epsffile{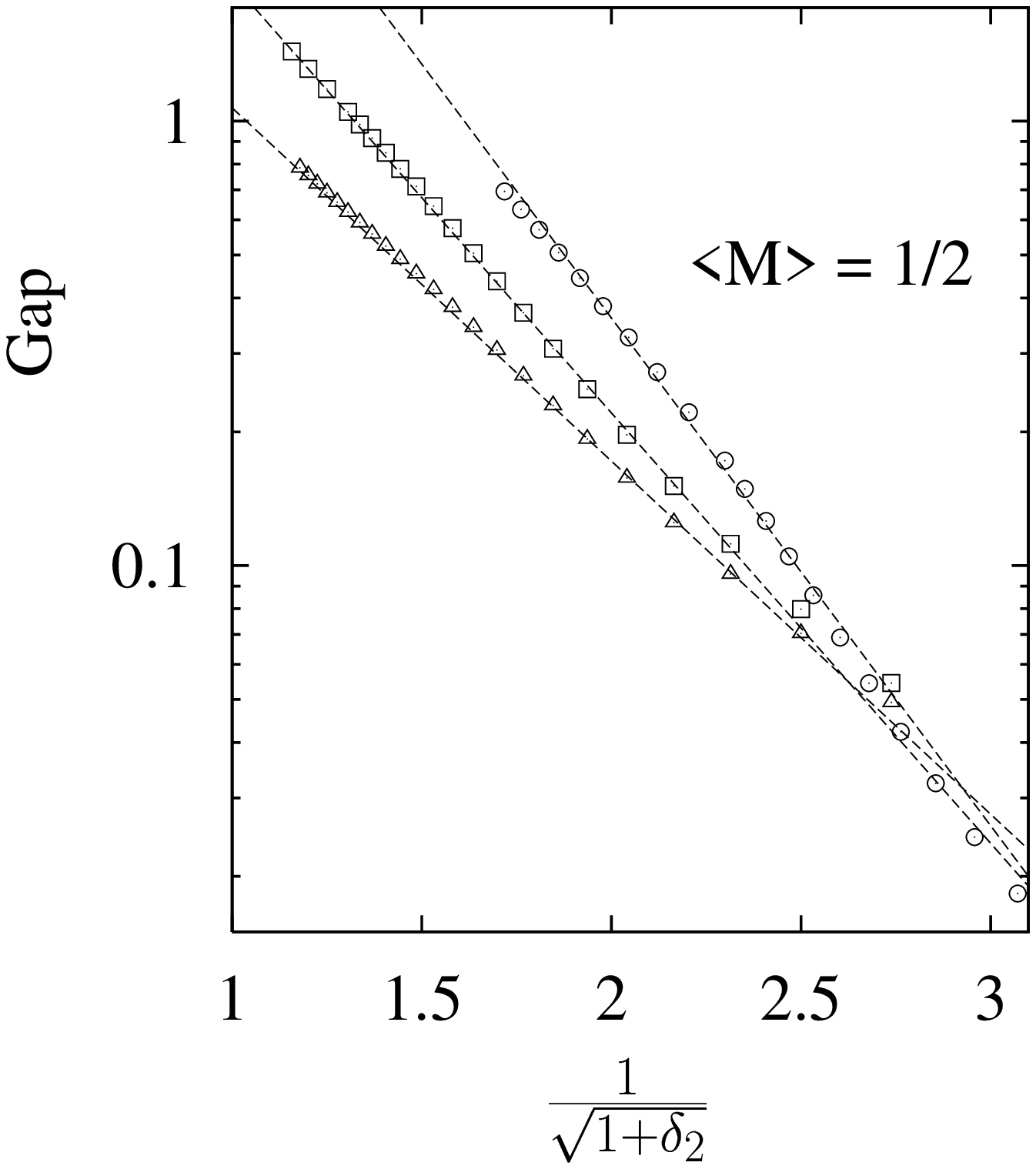}}
\vspace{-5.4cm}
\caption{Gap width extrapolations for $\langle M \rangle = 1/2$ suggesting
a KT singularity $\propto e^{-1/\sqrt{1+\delta_2}}\,$
on approaching $\delta_2 = -1$.
Triangles, squares and circles denote respectively
weak, intermediate and strong coupling regimes displayed in turn
in Figs. 6 (a), 6(c) and  6(e). Straight lines are guides to the eye.}
\end{figure}


\begin{thebibliography}{99}
\bibitem{comment} Certainly, a minimum of two Lanczos vectors
will do, though at the expense of a significant amount
of CPU time.
\end{thebibliography}
\end{document}